# Marking the Graphene Era in Disseminating the Redefined SI

**Albert F. Rigosi**

Physical Measurement Laboratory, National Institute of Standards and Technology, Gaithersburg, MD 20899, United States of America

## 1. The Graphene Era Begins

The history of quantum Hall standards stretches several decades and mostly begins with the use of GaAs given that 2D electron systems exhibit interesting quantum phenomena [1-31]. At the end 2000s, research in 2D materials like graphene became prevalent [32-35]. The QHE was observed and quickly became accessible to metrologists. QHR devices were becoming graphene-based, with fabrications performed by chemical vapor deposition (CVD) [36], epitaxial growth [35, 37], and the exfoliation of graphite [38]. Given the many methods of available graphene synthesis, efforts to find an optimal synthesis method for metrological purposes were underway. Exfoliated graphene was widely known to exhibit the highest mobilities due to its pristine crystallinity. It was a primary initial candidate as far as metrological testing was concerned. It was found in Giesbers *et al*. that devices constructed from flakes of graphene had low breakdown currents relative to GaAs-based counterparts, with currents on the order of 1 μA being the maximum one could apply before observing a breakdown in the QHE [38].

Among the forms of synthesis, epitaxial graphene (EG) proved to be the most promising for metrological applications [39]. In their work, Tzalenchuk *et al*. fabricated EG devices and performed precision measurements, achieving quantization uncertainties of about 3 nΩ/Ω. An example of their measurements is shown in Fig. 3 of Ref. [39], where (a) is a demonstration of how viable the graphene-based device was to be a suitable QHR standard. It was the start of a new chapter for standards, but additional work was required to exceed the stringent temperature requirement of 300 mK and low current capability of 12 μA, relative to modern day capabilities [39]. EG was also synthesized on SiC via CVD in 2015, showing that graphene could output standards-quality resistance at temperatures up to 10 K [40].

## 2 Establishing Graphene as a Global Standard

EG has been used as part of the electrical resistance dissemination service in the United States since early 2017. Plenty of work on evaluating EG as a suitable replacement for GaAs had been done in the years prior, but implementation of graphene into the formal global calibration chain did not occur until 2017 [41-45]. These preceeding years before global implementation were dedicated to optimizing basic fabrication processes so that EG-based QHR devices could be accepted as the formal replacement for GaAs-based QHR devices globally. Once prototypes emerged, EG's performance far surpassed that of GaAs-based devices. A compilation of these efforts can be linked to multiple institutes [46-53]. EG-based devices suitable for metrology required high-quality EG to the point of substantial scalability (centimeter-scale) as well as stability of its electrical properties to have a long shelf life and end-user ease-of-use.

In one example of EG-based QHR development, devices had become compatible with a 5 T table-top cryocooler system [45, 50]. The advantage of such a system also includes the removal of the need for liquid helium, enabling continuous, year-round operation and access to the QHR. The $v = 2$ plateau is the primary way to disseminate the ohm, much like GaAs-based QHRs. In fact, EG provides this plateau in a very robust way and for a large range of magnetic fields because of Fermi level pinning from the covalently bonded layer directly beneath the graphene layer [46]. Janssen *et al.* first demonstrated this type of measurement with a table-top system in 2015, pushing the bounds of operability to new levels.

For the group at the National Institute of Standards and Technology (NIST), an EG QHR device was scaled to a 1 kΩ standard using a binary CCC and a DCC [45]. The uncertainties that were achieved with this equipment matched those obtained in GaAs-based QHR systems (i.e., on the unit order of nΩ/Ω). The results of some of these measurements are shown in Fig. 1 at three different currents. The limit of nearly 100 µA gives EG-based QHRs the edge over GaAs, especially since these measurements were performed at approximately 3 K. Another example of graphene being established as the new QHR standard comes from Lafont *et al.*, whose work hardened the global question of when graphene would surpass GaAs in various respects [40]. An example of this investigation is shown in Fig. 2, where the sample of choice was CVD graphene on SiC. Overall, with the success that came out of the EG-based QHR efforts, the chapter for using GaAs-based QHRs as the leading standard had come to a smooth close.

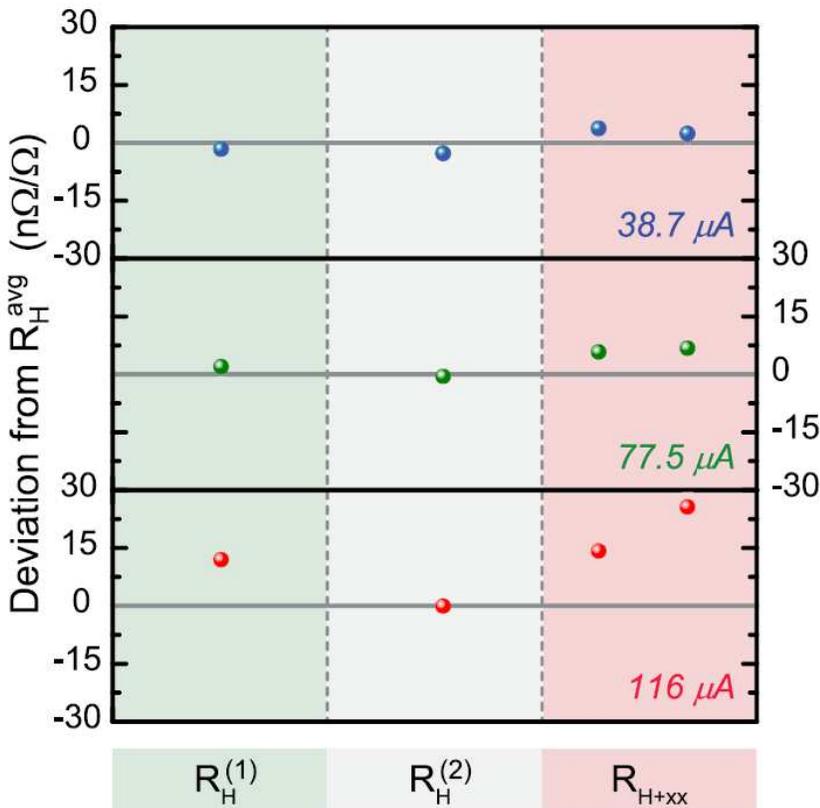

Figure 1. Binary winding CCC measurements were performed with an EG-based QHR device in a cryogen-free table-top system. The CCC measurement data are displayed in three rows, with the blue, green, and red data corresponding to the source-

drain currents of 38.7 μA, 77.5 μA, and 116 μA, respectively. Each column shows a comparison between the resistance of the ν = 2 plateau and a 1 kΩ resistor, with the left two representing orthogonal contact pairs ($R_H^{(1)}$ and $R_H^{(2)}$). The third column represents the average of two diagonal contact pairs formed by the latter two orthogonal pairs. It is labelled as $R_{K+xx}$ to give an idea of the impact of the longitudinal resistance. The type A measurement uncertainties are smaller than the data points and the type B uncertainties are under 2 nΩ/Ω. © 2019 IEEE. Reprinted, with permission, from [45].

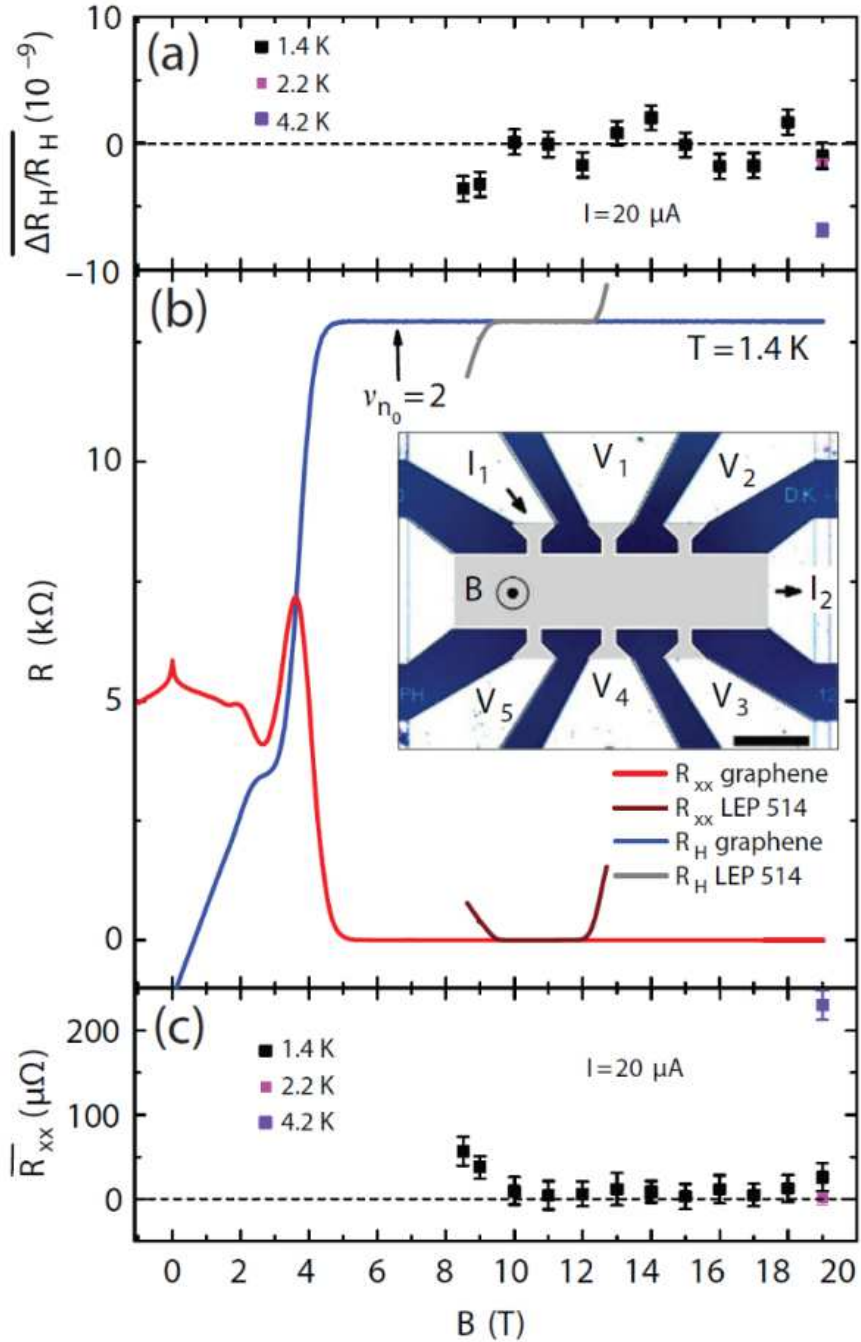

Figure 2. The magnetic field dependence for Hall resistance in CVD graphene is shown. (a) The Hall resistance ($R_H$) deviation is measured on the ν = 2 plateau at 1.4 K (black), 2.2 K (magenta) and 4.2 K (violet). (b) $R_{xx}$ and $R_H$ are measured and shown with an injection current of 100 nA, with both types of voltages measured using the ($V_1$, $V_2$) and ($V_2$, $V_3$) contact pads, respectively.

The magnetic field was energized to values between 1 T and 19 T for the graphene device (shown by the red and blue curves) and from 8 T to 13 T for the GaAs device (maroon and grey curves). The onset of the Landau level was calculated by using the carrier density at low magnetic fields, which is related to the slope of $R_H$ near zero field. The inset shows an optical image of the device (scale bar is 100 mm). (c) Data for precision measurements of $R_{xx}$ are shown at 1.4 K (black), 2.2 K (magenta) and 4.2 K (violet). The error bars represent combined standard uncertainties (1σ). Ref. [40] is an open access article distributed under the terms of the Creative Commons CC BY license, which permits unrestricted use, distribution, and reproduction in any medium.

## 3 Improvements in Measurement Infrastructure

Given the establishment of better QHR devices for resistance standards, the corresponding equipment and infrastructure with which one can disseminate the ohm had to also improve or at least be shown to maintain its compatibility with EG-based QHRs. The measurement of the ratio between a QHR device and a standard resistor must achieve uncertainties that are comparable to or better than the stability of those resistors. This criterion would support the notion of reliable traceability in that system. Measurements with a longer integration time produce better results because the ratios are well-maintained. The obvious benefit from room-temperature bridge systems like the DCC includes an ability to deliver measurements more frequently and year-round due to their operation not requiring cryogens [54]. Additionally, both DCCs and CCCs have been improved so that they are more user-friendly and automated, removing, at least in the case of the DCC, most of the dependence on specialized knowledge that one normally expects for the more complex cryogenic counterparts [16, 55]. Some NMIs have shown that the automated binary-ratio CCCs have an incredibly exceptional ratio accuracy with improved type B uncertainties, a useful feature when dealing with more precise ratios in QHR comparisons [56-58]. CCC scaling methods can achieve large resistance ratios of 100 or more, whereas DCCs trade off the required cryogens for having smaller ratios and lower current sensitivity.

The DCC in particular offers an opportunity to simplify measurement systems. The work by MacMartin and Kusters shows how they built a DCC for comparing four-terminal resistances [54]. It works by measuring a current ratio that corresponds to a particular voltage drop equality. That current ratio, represented as a turns ratio, was automatically maintained by a self-balancing DCC capable of being adjustable in part-per-million steps. The two current sources in the DCC are isolated so that there is no current in the potential circuit when fully balanced. Their DCC was able to measure and compare the ratio of two isolated direct currents accurately. A detailed description is laid out along with a discussion of its accuracy limitations [54]. In the optimum operating range, the authors achieved accuracies that were better than one part per million. The bridge was designed to accommodate the scaling of resistance standards from 100 Ω to a less than 1 mΩ and can be used for any ratio from 1 to 1000, thus permitting one to calibrate low value shunts whose accuracies would be limited only by the level of noise and the stability of the resistor [54].

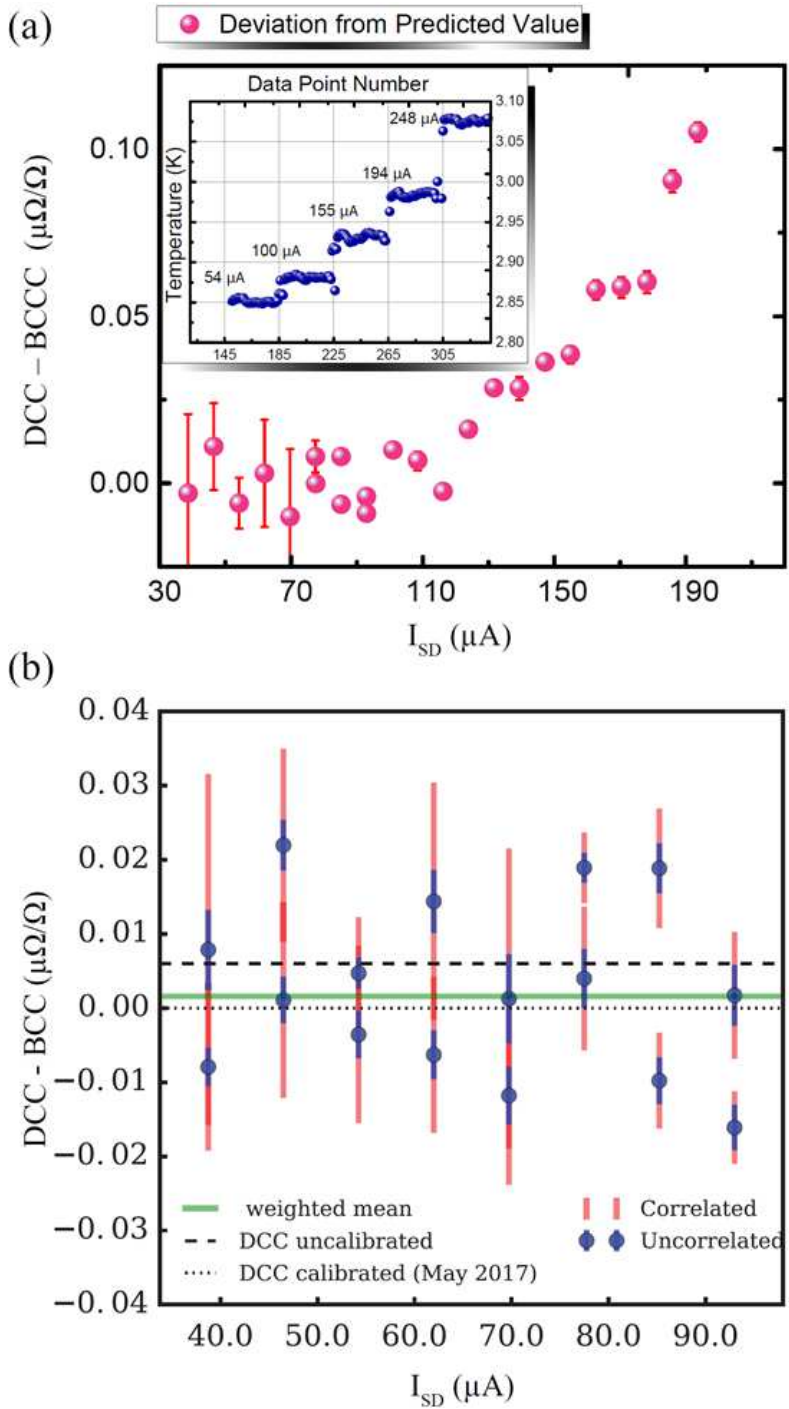

Figure 3. (a) Current dependence data of DCC ratios for a 1 kΩ resistor are shown in magenta, based on the same EG QHR device used with a binary winding CCC. The DCC results confirm that the QHR device remains quantized to within 0.01 µΩ/Ω up to approximately 116 µA. Data for heating to the sample temperature from current dissipation are shown in the inset with blue data points. Some type A measurement uncertainties (in red) are smaller than the data points. (b) DCC data for the ratio of the QHR to 1 kΩ are shown for increasing source-drain currents and are normalized to the average results of BCCC scaling (using 38.7 µA and 77.5 µA). The red error bars show the expanded ($k = 2$) correlated uncertainties whereas the blue error bars show the expanded ($k = 2$) uncertainty reported by the measurement device. © 2019 IEEE. Reprinted, with permission, from [45].

From the previously mentioned work at NIST [45], the two bridges are compared to demonstrate the applicability of using a room-temperature DCC with an EG-based QHR device. As shown in Fig. 3 (a), the DCC measurements extend the range of the applied source-drain currents when scaled from the EG-based QHR device to a 1 kΩ resistor. The deviations from the low-current binary winding CCC result exceed about 0.01 μΩ/Ω once the source-drain current exceeds 116 μA. The inset of Fig. 3 (a) shows how sample heating from high applied current affects the overall sample temperature. This information was relevant because high currents applied to EG-based QHR devices makes them more likely to operate well with the DCC, but reaching too high a temperature will begin to degrade the quality of the QHE. It was shown that temperatures below 4 K were still accessible despite applying nearly 250 μA to the device. Ultimately, the work found that table-top systems may limit how much current one may apply since currents as high as 550 μA would bring such a system to about 4 K, just shy of the boiling point of liquid helium [45].

As of the present day, CCC bridges outperform DCCs in terms of achievable uncertainties. However, in order to more easily disseminate QHR technology globally, room temperature DCCs are preferred in terms of ease-of-use and fewer resource needs, like not needing cryogens to operate. For this reason, a comparison between the two methods was examined and is shown in Fig. 3 (b), where the number of DCC data points averaged was varied inversely with the square of the applied voltage $V \approx I_{SD} \times R_H$ to obtain a similar type A uncertainty for all values of the source-drain current, with measurement durations ranging from 110 min for the 0.5 V measurements to 24 min for the 1.2 V measurements. Further, in Fig. 3 (b), the uncertainties indicated in blue were produced by the DCC software. The larger, red error bars are uncertainties that take into account any statistical correlations in the data, as thought out by other research efforts [59]. Potential noise from the cryogen-free mechanical refrigeration system may also interfere to some degree with the balancing algorithm.

Even though DCCs would make global dissemination easier for institutes external to NMIs, CCC technology is still important for metrologists. Williams *et al*. demonstrated a design for an automated CCC in order to perform routine NMI measurements [58]. Their system uses a type II CCC for use in a low loss liquid helium storage vessel and may be continuously operated with isolated supplies coming from the mains power. All parameters were shown to be digitally controlled, and the noise sources present in the system were analysed using the standard Allan deviation, leading to the conclusion that one may eliminate non-white noise sources simply by choosing the appropriate current reversal rate.

As we move forward, NMIs have both anticipated, and in many cases achieved, these goals for EG-based QHR devices. Lastly, to begin expanding on the functionality of EG-based devices, it would need to be shown that precise resistance scaling could be done to reference resistors from the QHR by using voltages large than those available in standard DCCs and CCCs. Further improvements to this graphene technology, as well as resistance standards in general, include the implementation of superconducting contacts, p-n junctions, and topological insulators that exhibit the quantum anomalous Hall effect [60-84]. These efforts are underway and hope to eventually accelerate dissemination efforts.

## Acknowledgements

The authors wish to acknowledge S. Mhatre, A. Levy, G. Fitzpatrick, and E. Benck for their efforts and assistance during the internal review process at NIST. Commercial equipment, instruments, and materials are identified in this paper in order to specify the experimental procedure adequately. Such identification is not intended to imply recommendation or endorsement by the